\documentclass[aip, jcp, twocolumn, superscriptaddress, showkeys, 10pt]{revtex4-1}
\usepackage[hidelinks]{hyperref}
\pdfoutput=1
\usepackage[english]{babel}
\usepackage{graphicx}
\usepackage{color}
\usepackage{amssymb, amsfonts, amsmath}

\usepackage[caption=false]{subfig}
\addto\extrasenglish{%
}
\makeatletter
\let\oldtheequation\theequation
\renewcommand\tagform@[1]{\maketag@@@{\ignorespaces#1\unskip\@@italiccorr}}
\renewcommand\theequation{(\oldtheequation)}
\makeatother

\begin{document}

%%%%%%%%%%%%%%%%%%%%%%%%%%%%%%
\title{Estimation of the critical behavior in an active colloidal system with 
Vicsek-like interactions}
\date{\today}
\author{Benjamin Trefz}
\affiliation{Johannes Gutenberg University Mainz, Department of Physics, Staudingerweg 7, 55128 Mainz, Germany}
\affiliation{Graduate School Material Science in Mainz, Staudinger Weg 9, 55128 Mainz, Germany}
\author{Jonathan Tammo Siebert}
\affiliation{Johannes Gutenberg University Mainz, Department of Physics, Staudingerweg 7, 55128 Mainz, Germany}
\author{Thomas Speck}
\affiliation{Johannes Gutenberg University Mainz, Department of Physics, Staudingerweg 7, 55128 Mainz, Germany}
\author{Kurt Binder}
\affiliation{Johannes Gutenberg University Mainz, Department of Physics, Staudingerweg 7, 55128 Mainz, Germany}
\author{Peter Virnau}
\affiliation{Johannes Gutenberg University Mainz, Department of Physics, Staudingerweg 7, 55128 Mainz, Germany}

\begin{abstract} 
We study numerically the critical behavior of a modified, active Asakura-Oosawa 
model for colloid-polymer mixtures. The colloids are modeled as self-propelled
particles with Vicsek-like interactions. This system undergoes phase separation 
between a colloid-rich and a polymer-rich phase, whereby the phase diagram 
depends on the strength of the Vicsek-like interactions. Employing a subsystem-block-density 
distribution analysis, we determine the critical point and make an attempt
to estimate the critical exponents. 
In contrast to the passive model, we find that the critical point is not located 
on the rectilinear diameter.
A first estimate of the critical exponents $\beta$ and $\nu$ is 
consistent with the underlying 3d-Ising universality class observed for the 
passive model.
\end{abstract}


\maketitle 

\section{Introduction}

Active particles are intrinsically non-equilibrium systems which have some means 
of self-propulsion. 
This can be a motor or flagellum, but can also be induced by the solvent and/or
external sources. 
%, e.g. via vibrations \cite{narayan2006nonequilibrium, deseigne2010collective} or light \cite{ibele2009schooling}. 
In all cases, some form of energy is converted into kinetic energy that results 
in the self-propulsion.  
This general definition encompasses a large variety of systems on different
scales. % that are considered active.
Besides rather large macroscopic systems such as flock of birds or school of 
fish\cite{ballerini2008interaction, katz2011inferring}, 
active particles are also found on a micrometer scale. 
Such systems include actin filaments \cite{schaller2010polar}, 
and microtubules \cite{sumino2012large, sanchez12spontaneous} that are moved by 
motor proteins in a plane and can be observed via microscopes.
Some bacteria are able to propel themselves and can show density-dependent 
phase separation \cite{cates2010arrested, zhang2010collective}.
Sperm cells cooperate due to hydrodynamic interactions and form clusters 
\cite{Yang2008}. 
It is even possible to alter microorganisms and make them thereby active, e.g. 
by attaching an artificial, magnetically activated flagellum 
\cite{dreyfus2005microscopic}. Another approach is to combine an already 
self-propelled particle, e.g. a sperm cell, with an externally controllable 
non-motile part, e.g. a magnetic microtube \cite{Magdanz2013}.

In soft matter systems, colloids play an important role as a model system since 
they provide an ideal environment to compare experiment, computer simulation,
and theory. In particular, interactions between colloidal particles are 
tunable, and one can follow the motion of individual colloids by confocal video
microscopy techniques.
Active particles are no exception and a variety of systems with 
self-propelled particles have been studied. 
%Colloidal systems that are driven from the outside, e.g. by shining light on the 
%solution, allow for direct comparison of the active and the passive system 
%\cite{jiang2010active, buttinoni2012active, palacci2013living}.
%Another approach is to provide the ``fuel'' necessary for the self-propulsion in 
%the solution \cite{howse2007self, ke2010motion, palacci2010sedimentation}. 
%In these examples, the solution contains hydrogen peroxide and the active 
%constituents are so-called Janus-type particles, where one hemisphere is coated 
%with platinum. The conducting hemisphere acts as catalyst for the reaction of 
%hydrogen peroxide to water and oxygen and thus ``consumes'' the fuel, 
%which in turn propels the particle forward.
Colloidal systems that are driven from the outside allow for direct comparison 
of the active and the passive system. Self-propulsion can be
achieved in many ways, e.g. thermophoresis, diffusiophoresis, or electrophoresis. 
Thermophoresis can be realized via an external light source that heats the 
sample generating a temperature gradient \cite{jiang2010active}.
Self-diffusiophoresis has been observed in a binary, near-critical solvent 
\cite{buttinoni2012active}. Other swimmers exploit a chemical reaction to 
maintain a local gradient
%the ``fuel'' necessary for the self-propulsion is provided in the solution 
\cite{howse2007self, ke2010motion, palacci2010sedimentation, palacci2013living}. 
In these examples, the solution contains hydrogen peroxide and the active 
constituents are so-called Janus-type particles, where one hemisphere is coated 
with platinum. The conducting hemisphere acts as catalyst for the reaction of 
hydrogen peroxide to water and oxygen and thus ``consumes'' the fuel, 
which in turn propels the particle forward.
Self-propulsion induced by electric fields can be realized via Quincke rotation
of the colloid \cite{bartolo13emergence} or via a metallic Janus-type 
particles \cite{jing16reconfiguring}.

In the last years several different models to study active particles have been
discussed \cite{grossman2008emergence, bialke2012crystallization, 
farrell2012pattern, Das2014, bialke2015active, cates2015motility, Trefz2016}.
%In active systems phase transitions are of special interest, as a lot of 
Many active systems show the tendency to form clusters, e.g., flock of birds, 
school of fish, and colony of bacteria. This raises the question of phase 
separation, which has been analyzed in various numerical investigations 
\cite{redner13structure, Das2014, stenhammar2014phase, wysocki14cooperative, 
bialke2015active, prymidis15selfassembly, Trefz2016}.
Studies with active particles often consist of active and passive particles,
e.g., motile bacteria in a polymer background 
\cite{schwarz2012phase, stenhammar15activity, grosberg15activity}.
The model we study here is a variation of the well known Asakura-Oosawa (AO) model 
\cite{asakura1954,asakura1958,vrij1976,vink2005,zausch2009statics,binder2014},
which consists of two particle types, colloids and polymers. 
In our active model, the colloidal particles become self-propelled with 
Vicsek-like interactions\cite{Das2014, Trefz2016}. 
This facilitates the phase separation compared to the passive model, 
since the Vicsek-like activity induces the formation of clusters. 
This result, together with the static and dynamic behavior of this model, has 
been reported in \cite{Das2014, Trefz2016}. 
Importantly, the passive AO model already features a phase transition and 
belongs to the Ising universality class.
From the active model, the passive model is recovered in the limit 
when the strength of the self-propulsion goes to zero. 
%Thus, for small activity the critical point will be preserved in 
%this model. 

%The determination of the critical point in a non-equilibrium system is quite 
%difficult. 
A very interesting question is whether critical phenomena in a 
non-equilibrium system belong to a known universality class of some related 
equilibrium system or form a new class. A first step to address this issue, 
of course, is to find the location of the critical point.
For systems under shear \cite{winter2010determination} one finds that the system 
changes its behavior towards the mean-field universality class in the limit of 
infinite shear \cite{hucht2009nonequilibrium}. 
%\cite{onuki1979nonequilibrium, beysens1983}.
%For active particles, %to our knowledge, only in 
The critical point of the classical Vicsek model 
%the critical point 
has been determined successfully
\cite{Vicsek1995, czirok1997spontaneously}. 
It should be noted that this was a heavily discussed issue, and the continuous
phase transition from disordered to ordered state was questioned  
\cite{gregoire2004onset}. 
It was later shown that the way noise is introduced in the classical Vicsek 
model can change the order of the phase transition
\cite{nagy2007new, aldana2007phase}.
For the determination of the critical point numerous simulations at different 
system sizes, densities and noise values had to be performed.
A different approach has recently been proposed \cite{prymidis2016}, where the 
critical point of an active Lennard-Jones system has been determined by fitting
various power laws and assuming an exponential dependence on parameters.
In other models, the critical point in the active case turned out to be at 
infinite density and could thus not be determined directly via simulation 
\cite{solon2015flocking}. 

In this paper, we will discuss the determination of the critical point in a
system of self-propelled particles using the subsystem-block-distribution 
analysis. The method will be general and thus should apply to any density-driven 
phase separation that features a second order phase transition. 
Exemplary we will determine the 
critical point of our active Asakura-Oosawa model.  

\section{Model and Methods}
\begin{figure}[htb]
    \centering
    \def\svgwidth{0.85\linewidth}
    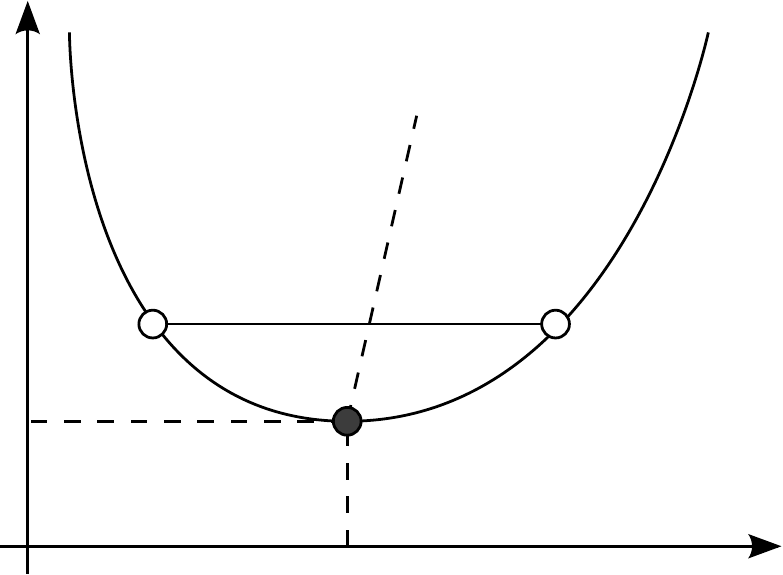\\[0.5cm]
    \def\svgwidth{0.85\linewidth}
    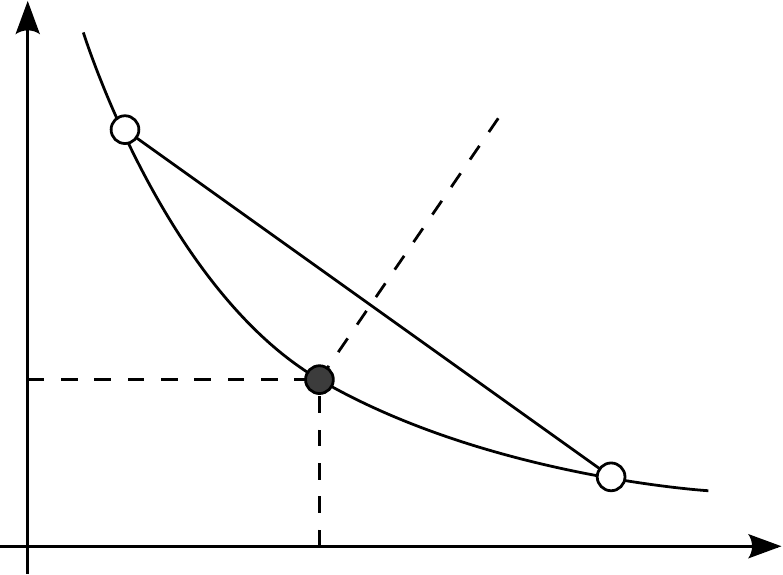\\[0.5cm]
    \def\svgwidth{0.85\linewidth}
    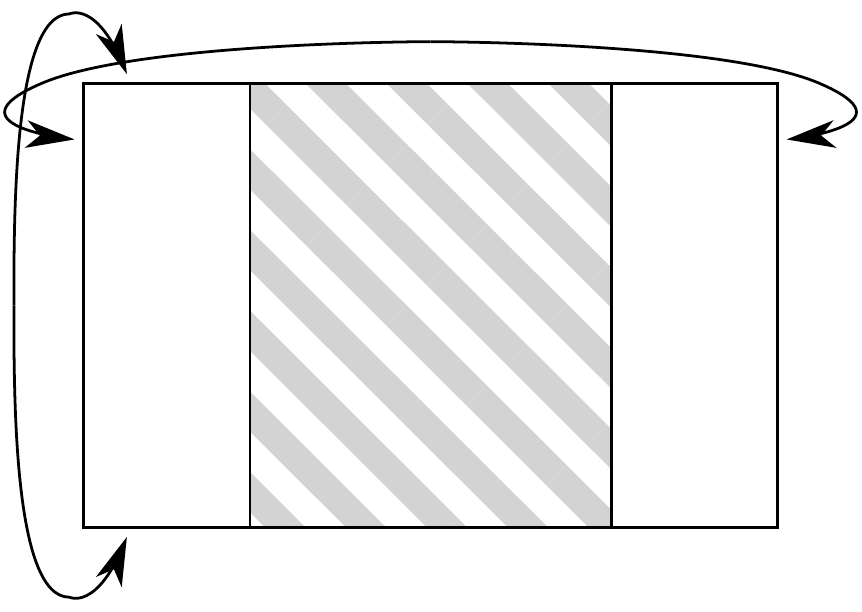
    \caption{%
    Schematic drawing of (a) the phase diagram for the colloid
    packing fraction $\eta_{\text{col}}$ plotted against the intensive 
    thermodynamic variable $\eta_{p}^{r}$, (b) the phase diagram for two 
    extensive variables $\eta_{\text{col}}$ and $\eta_{\text{pol}}$. 
    (c) shows a schematic representation of a system in the two phase region. 
    The two tie points from (b) can be obtained by extracting the packing 
    fractions in the liquid and the gas phases. With this recipe the coexistence 
    curve in \autoref{Fig2-phasediagram} has been determined.}
    \label{Fig-schematic}
\end{figure}
Let us first recall why the location of the critical point of a non-equilibrium 
system is much more difficult to find than for an equilibrium system. In the 
latter, we can study the phase behavior choosing an intensive thermodynamic 
variable as a control variable; e.g. in a colloid-polymer mixture the chemical
potential of the polymers (or a related variable, such as the so-called 
polymer reservoir packing fraction $\eta_{p}^{r}$) are commonly used. In the 
resulting phase diagram, the critical point then occurs at the minimum of the 
coexistence curve, and the tie lines connecting coexisting vapor-like and 
liquid-like phases of the colloidal suspension are horizontal lines
(\autoref{Fig-schematic}(a)). In thermal equilibrium, the thermodynamic 
relations allow the translation of this phase diagram in a representation with 
two densities of extensive variables, the colloid packing fraction 
$\eta_{\text{col}}$ and the polymer packing fraction $\eta_{\text{pol}}$   
(\autoref{Fig-schematic}(b)). Then the tie lines no longer are horizontal lines but 
rather are oriented under an a priori unknown angle, and the critical point is
not on a straightforwardly defined position on the coexistence curve, but rather
nontrivial to find in this statistical ensemble.

In the non-equilibrium system containing active colloids, intensive 
thermodynamic variables no longer are well-defined, in contrast to extensive 
variables (number of colloids $N_{\text{col}}$ and number of polymers 
$N_{\text{pol}}$ in the considered volume) and their densities, which are still well 
defined. We ask whether phase separation in
a gas-like and liquid-like phase also occurs, and if so, estimate the 
corresponding phase diagram. This task was already attempted in Refs.
\citenum{Das2014, Trefz2016}, looking for phase coexistence in simulation volumes
elongated in $z$-direction, where in the two phase region a liquid domain 
separated by two (on average planar) interfaces from the gas occurs 
(\autoref{Fig-schematic}(c)). This means that the local densities of colloids 
and polymers separate in gas and liquid domains, and the end points of the tie 
line in \autoref{Fig-schematic}(b) can be found, but only for states far away 
from the critical point. For finding the location of the critical point, 
obviously a different approach must be sought, since near criticality the 
density differences between the coexisting phases are small. Strong and 
long-lived density fluctuations occur, and the interfaces 
become very rough and diffuse. The same difficulty would occur if we would use
$\eta_{\text{col}}$, $\eta_{\text{pol}}$ as variables in a simulation of a 
colloid-polymer mixture in equilibrium, but there \autoref{Fig-schematic}(a) 
provides for a more convenient alternative, e.g. one records the probability 
distribution $P(\eta_{\text{col}})$ at fixed $\eta_{p}^{r}$, finding the 
end-points of the tie line in \autoref{Fig-schematic}(a) from the peaks of that 
distribution, and analyzing the merging of the peaks near criticality in terms
of a finite size scaling analysis \cite{vink2005}.

In the following, we study a model of Vicsek-like interactions between active particles.
A detailed description of the model can be found in Refs. \citenum{Das2014, Trefz2016}.
The binary system is a variant of the well-known Asakura-Oosawa (AO) model and 
consists of colloids (c) and polymers (p) \cite{zausch2009statics}.
The potentials are given by:
\begin{align}
    U_{cc}(r) &= 4 \epsilon_{cc} \left[ \left(\dfrac{\sigma_{cc}}{r} \right)^{12} 
			     - \left(\dfrac{\sigma_{cc}}{r} \right)^{ 6} 
			     + \dfrac{1}{4} \right]\label{eq:potential_cc}\\
    U_{cp}(r) &= 4 \epsilon_{cp} \left[ \left(\dfrac{\sigma_{cp}}{r} \right)^{12} 
			     - \left(\dfrac{\sigma_{cp}}{r} \right)^{ 6} 
			     + \dfrac{1}{4} 
		       \right] \label{eq:potential_cp}\\     
    U_{pp}(r) &= {\scriptstyle 8 \epsilon_{pp} \left[  1 
			     - 10 \left(\dfrac{r}{r_{pp}} \right)^{3} 
			     + 15 \left(\dfrac{r}{r_{pp}} \right)^{4} 
			     -  6 \left(\dfrac{r}{r_{pp}} \right)^{5} 
		       \right]} \quad , \label{eq:potential_pp}
\end{align}
if $r$ is smaller than the respective cut-off radius $r_{cc} = 2^{1/6} 
\sigma_{cc}$, $r_{cp} = 2^{1/6} \sigma_{cp}, r_{pp} = 2^{1/6} \sigma_{pp}$,
and zero otherwise. The other parameters are chosen according to 
\cite{zausch2009statics, Das2014, Trefz2016}
$\epsilon_{cc} = \epsilon_{cp} = 1$, 
$\epsilon_{pp} = 0{.}0625$, $\sigma_{cc} = 1$, $\sigma_{cp} = 0.9$, and 
$\sigma_{pp} = 0.8$. To be consistent with the literature, we calculate 
the packing fractions $\eta_{\alpha}$ as
$\eta_{\alpha} = \rho_{\alpha} V_{\alpha}$, where 
$V_{\alpha} = \pi d_{\alpha}^{3} / 6$ is the volume of a single sphere and 
$d_{\alpha}$ is the Barker-Henderson diameter \cite{barker-henderson} of the 
colloids or in case of the polymers $0.8 d_{cc}$. To thermostat the system we 
use a Langevin thermostat in our MD simulation \cite{grest1986, frenkelsmit}.

The equations of motion are
\begin{align}
m \ddot{\vec{r}}_{i} = - \vec{\nabla} U - \gamma m \dot{\vec{r}}_{i} + \sqrt{2 \gamma k_{B} T m} \; \vec{R}_{i}(t) \quad,
\label{eq:langevin}
\end{align}
where $m = 1$ is the particle mass (for all particle types), 
$\gamma = 1$ is the friction coefficient, 
$U$ is the interparticle potential, 
$T = 1$ is the temperature, and 
$\vec{R}$ is a zero-mean unit-variance Gaussian white noise. 
We use a Velocity Verlet algorithm with a time step $\Delta t = 0{.}002 t_{0}$, 
with $t_{0} = \sqrt{\sigma_{cc}^{2} m / \epsilon_{cc}}$.

\begin{figure}[htb]
\centering
\includegraphics[]{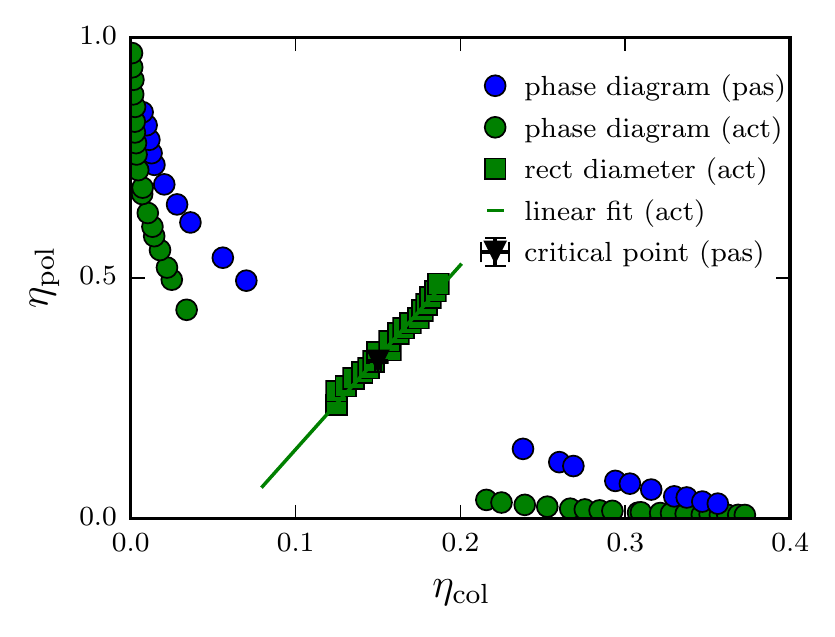}
\caption{%
Phase diagram of the active and passive system together with a fit of the 
rectilinear diameter of the active system (\autoref{eq:rectilinear_dia_cAO}).
The critical point of the passive system is taken from \cite{zausch2009statics}.
}
\label{Fig2-phasediagram}
\end{figure}

In the active version we employ a variation of the Vicsek 
model~\cite{Das2014, Trefz2016} on top of the passive AO model for the colloids. 
For that we still solve the Langevin equation first just as in
the passive model. 
The resulting velocity is then modified by an additional force  
\begin{align}
\vec{f}_{i} &= f_{A} \cdot \dfrac{ \left< \vec{v}_{j} \right>_{R} }{ \left< |\vec{v}_{j}| \right>_{R}}
\label{eq:vicsek}
\end{align}
acting on particle $i$.
The constant force is set to $f_{A} = 0$ for the polymers and 
$f_{A} = 10$ for the colloids in this work. 
The brackets $< >_{R}$ denote an average over all colloids in a sphere of radius $R$, with 
$R = \sqrt{2} \; r_{cc}$ being the cut-off radius for what is considered a neighbor.
In the active model we observe enhanced phase separation, as can be seen in 
\autoref{Fig2-phasediagram}, and which was already discussed in 
Refs. \citenum{Das2014, Trefz2016}.

This system is out of equilibrium, and the temperature $T$ used in 
\autoref{eq:langevin} does not characterize fluctuations of velocity or other 
variables in the system as demonstrated in earlier work 
\cite{Das2014, Trefz2016}.
As discussed in the introduction, the distribution function 
$P(\eta_{\text{col}}, \eta_{\text{pol}})$ is the quantity that contains the 
desired information on phase separation (and associated criticality) in the 
system. However, in our system (we choose a cubic box of linear dimension $S$
with periodic boundary conditions throughout, containing $N_{\text{col}}$ 
colloids and $N_{\text{pol}}$ polymers) both $\eta_{\text{col}}$ and 
$\eta_{\text{pol}}$ are fixed, and hence the distribution function of the total
system is meaningless. However, a way out of this dilemma is the application of
the so-called subsystem-block-density distribution 
\cite{binder1981finite, rovere1988block, rovere1993simulation, watanabe2012phase} 
which we will refer to as subbox method. Here, a big canonical simulation box is
simulated and divided into many smaller subboxes. In these subboxes the particle
number is allowed to fluctuate, thus a ``quasi'' grand canonical system is 
simulated. For each subbox we can then determine the higher moments of the 
density distribution
\begin{align}
    m^{2}(\alpha)        &= \frac{1}{N^{3}} \sum \limits_{i} \left(\rho_i(\alpha) - \bar{\rho}(\alpha)\right)^{2}\label{eq:m2} \\
    m^{4}(\alpha)        &= \frac{1}{N^{3}} \sum \limits_{i} \left(\rho_i(\alpha) - \bar{\rho}(\alpha)\right)^{4}\label{eq:m4} \\
\intertext{and calculate the cumulant as}
    U_{N}(\eta_{\alpha}) &= \dfrac{\langle m(\alpha)^{4} \rangle}{\langle m(\alpha)^{2} \rangle^{2}} \label{eq:cumulant} \quad , 
\end{align}
where $\alpha$ is either col or pol, $\rho_{i}(\alpha)$ is the 
density of particle $\alpha$ in subbox $i$, $\bar{\rho}(\alpha)$ 
is the average density of particles of type $\alpha$ in the system, and $N^3$ 
is the total number of subboxes of the system. 
Note that in \autoref{eq:cumulant} the average
$\langle...\rangle$ indicates an average over multiple, independent simulation 
snapshots, while $m^{2}$ and $m^{4}$ are already averaged over all subboxes 
of the same size.
With this method the same $N_{\text{col}}N_{\text{pol}}$VT trajectory can be 
used to compute all subbox systems simultaneously. 
This reduces the computation time substantially, although care is required, 
since the fluctuations observed for different subbox sizes $L$ clearly are not 
uncorrelated.

For the analysis one has to select proper subbox sizes. 
In Ref. \citenum{rovere1993simulation} the authors estimate that the subbox size $L$
should be chosen in a way that $\xi \ll L \ll S$, where $\xi$ is the correlation
length and thus a priori unknown but constant. 
Unfortunately, there is no obvious way to choose the optimal subbox sizes.  
Surely, the resulting subbox volume should not be too small, 
since the fluctuation of the density, corresponding to the addition or 
subtraction of a single particle, is getting bigger. 
Hence, the studied distribution would change from Gaussian to Poissonian. 
On the other hand the subbox should not be too big, as then there are too few 
subboxes and the correlation between them is increasing. 
Therefore, the overall explored phase space gets to narrow and thus the 
systematic errors due to the finite size of $S$ become too large 
(the finite size analysis \cite{binder1981finite, rovere1988block, 
rovere1993simulation} ignores the presence of a further non-zero 
scaling variable $L/S$ completely!).  
Since both effects are difficult to quantify, 
we choose the subbox sizes empirically, by only using subboxes that 
show a reasonable behavior far from the critical point.

\section{Results}
\subsection{Rectilinear diameter}\label{secA}
In order to determine the critical point of the active ($f_{A} = 10$) system we 
use an iterative approach. First we assume that the law of rectilinear diameter,
which the passive system approximately follows, is still true in the non-equilibrium model and
we can thus write
\begin{align}
    \frac{1}{2} \cdot (\eta_{\text{pol}}^{\text{gas}} + \eta_{\text{pol}}^{\text{liquid}}) = 
    \frac{a}{2} \cdot (\eta_{\text{col}}^{\text{gas}} + \eta_{\text{col}}^{\text{liquid}}) + b \quad.
    \label{eq:rectilinear_dia_cAO}
\end{align}
The rectilinear diameter for the active system is shown
in \autoref{Fig2-phasediagram} as green squares and the green line represents the fit to 
\autoref{eq:rectilinear_dia_cAO}, where $a=3.83$ and $b=-0.24$ are the 
resulting fit parameters.
\begin{figure}[hbt!]
 \centering
 \subfloat{\includegraphics{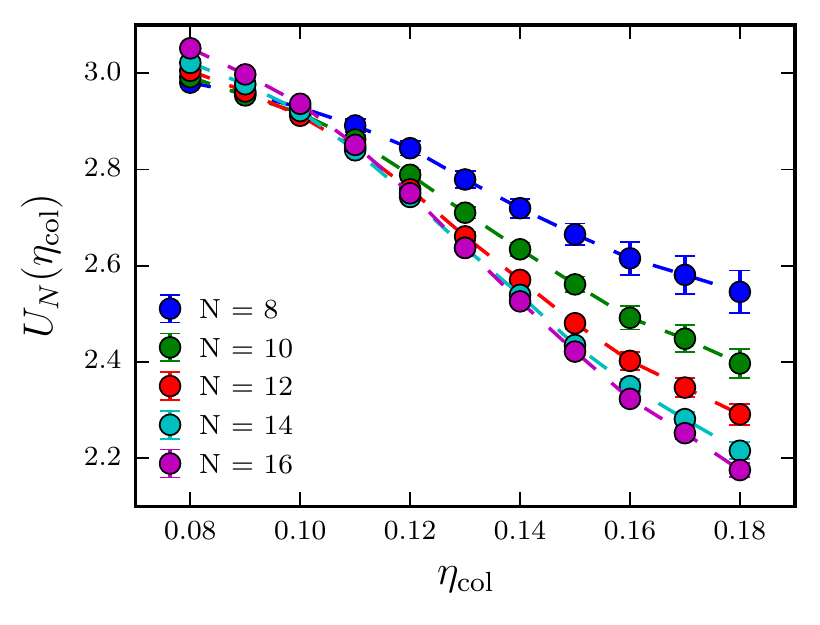}}\\\vspace{-0.9cm}
 \subfloat{\includegraphics{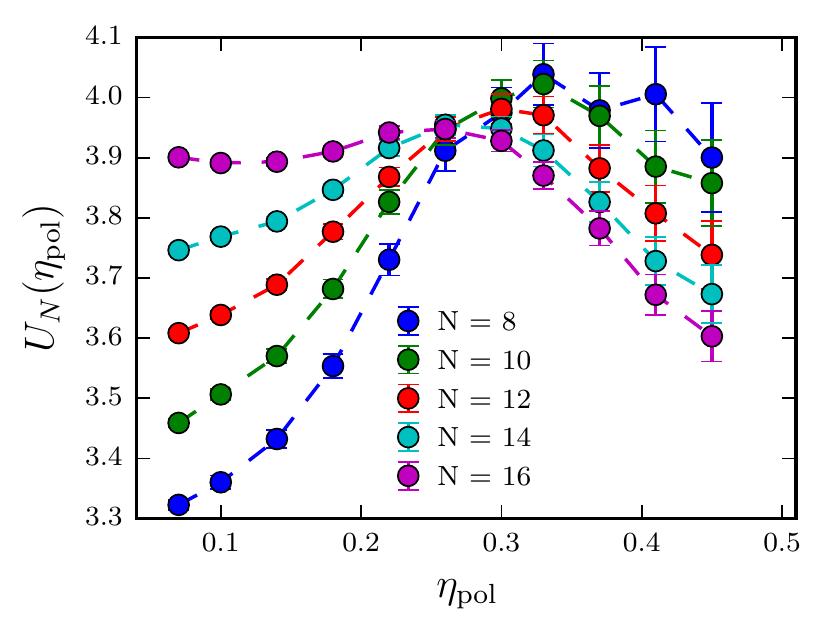}}\vspace{-0.5cm}
 \caption{%
 Crossing of the Binder cumulants along state points on the fitted rectilinear 
 diameter from \autoref{Fig2-phasediagram}. (a) The colloid cumulant 
 $U_{N}(\eta_{\text{col}})$ is plotted against the colloid packing fraction 
 $\eta_{\text{col}}$. The intersection point is read off as 
 $\eta_{\text{col}}^{\text{crit}} = 0.103(5)$. (b) The same state points are 
 analysed for the polymers. The intersection is at state points with higher 
 densities than for the colloids, thus the statistical error of this 
 intersection is larger. The critical polymer packing fraction is estimated as
 $\eta_{\text{pol}}^{\text{crit}} = 0.278(8)$.
 }
 \label{Fig3-rectilineardiameter}
\end{figure}
For the known region of the phase diagram, the active model seems to follow the 
law of rectilinear diameter and its difference to the passive system is minor, 
as the critical point of the passive system falls nicely onto the fit. 
However, it should be remembered that the ``law of rectilinear diameter'' 
\cite{rowlinson1982liquids} is not a general law of statistical thermodynamics, 
but rather can be derived only in the framework of mean field type theories. 
In fact, very close to the critical point deviations from this ``law'' are 
expected already for systems in thermal equilibrium 
\cite{sengers1980universality, kim2003asymetric, kim2003}, but for passive
systems deviations are typically negligible.
Therefore, we assume for now that the active system also follows the law of 
rectilinear diameter and simulate the active system for different 
state points along the green line in \autoref{Fig2-phasediagram}. 
We use a cubic simulation box with $S=48\sigma_{cc}$ and subdivide the system 
into many small subboxes $N=8,10,12,14,16$ to calculate the moments and 
cumulants as defined in Eqs.~\ref{eq:m2}-\ref{eq:cumulant}.
The length of each subbox is then $L=S/N$. 
In \autoref{Fig3-rectilineardiameter}, the intersection of the Binder cumulants 
$U_{N}$ are shown along the path of the rectilinear diameter. 
In the ideal case (where the limit $S \to \infty$ could be taken and $L$ is
extremely large), one would hope that the cumulants cross in a single 
($L$-independent) crossing point. In reality, this is not the case, the crossing
points are spread out over some region (this is expected due to the fact that
$L$ and $S$ are not large enough to reach the finite size scaling limit fully)
\cite{rovere1993simulation}.
However, from the multiple crossings one can use the average value as an 
estimate for the crossing point and the standard deviation as an estimate of error.
For the critical colloid packing fraction we find 
$\eta_{\text{crit}}^{\text{col}} = 0.103(5)$, while the critical polymer
packing fraction is determined as $\eta_{\text{crit}}^{\text{pol}} = 0.278(8)$.
While both cumulants $U_{N}(\eta_{\text{col}})$
and $U_{N}(\eta_{\text{pol}})$ cross for all subbox sizes analyzed, the crossing 
occurs at different state points.
Thus, the critical point will not fall
onto this line of rectilinear diameter but will be slightly shifted.
Note that a deviation from the law of rectilinear diameter has been observed
for a different active model as well \cite{prymidis2016}.
In an equilibrium system, it has been shown that the critical parameters can 
reasonably well be determined independently of each other \cite{rovere1993simulation}.
Therefore, we can interpret the cumulant intersections as an approximation of 
the critical point. 
In the following we will improve the accuracy with which the critical point 
is estimated by two independent approaches. 

\subsection{Extrapolation from the homogeneous region}
Due to the rather large value of the slope of the rectilinear diameter the colloid packing fraction has 
a better accuracy than the polymer packing fraction, which can also be seen in 
\autoref{Fig3-rectilineardiameter}. Therefore, we determine the critical polymer packing 
fraction by extrapolating the susceptibility from the homogeneous phase to the 
critical colloid packing fraction, which for now we assume to be correct. 
For that the order parameter susceptibility is determined from the two 
dimensional probability distribution $P(N_{\text{col}}, N_\text{{pol}})$ which, 
in the homogeneous phase, has contour lines that are ellipses.
As shown in Ref. \citenum{zausch2009statics}, the susceptibility is proportional to
\begin{align}
  \chi_{+} &\propto \dfrac{(\text{HWHM})^{2}}{N_{\text{col}}+N_{\text{pol}}} = W_{+} \quad.
  \label{eq:xi_plus}
\end{align}
Here HWHM stands for the half-width half-maximum of the 
long axis of $P(N_{\text{col}}, N_\text{{pol}})$, which is determined by fitting 
an ellipse to the distribution's equi-probability line at $0.5 P_{\text{max}}$.
\begin{figure}[!hbt]
 \centering
  \includegraphics{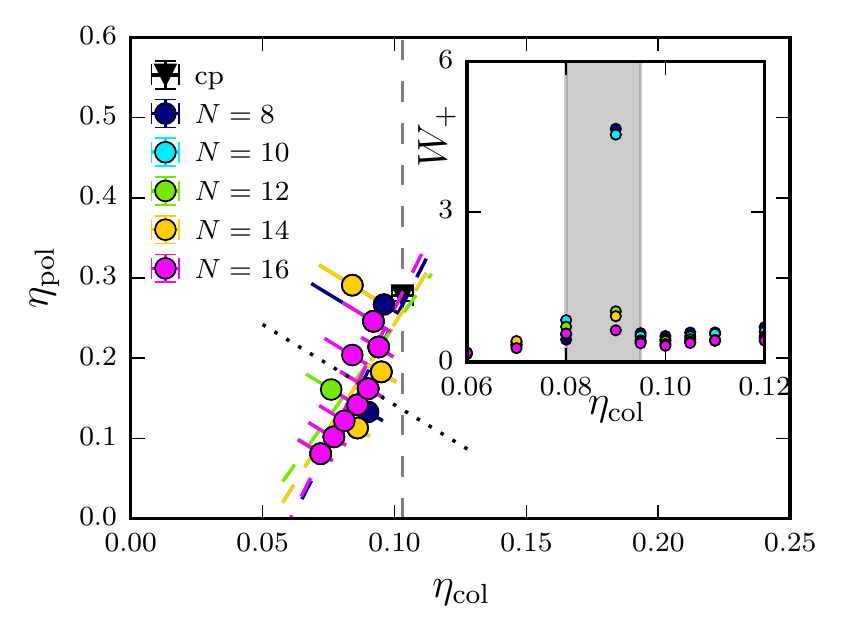}
 \caption{%
 Extrapolation of the maximum order parameter susceptibility $\chi_{+}$ 
 determined for individual slopes in the phase diagram. The error bars indicate
 the distance to the next simulated state point, and can thus be asymmetrical. 
 The locus of maximal susceptibility indicates a linear behavior over a long 
 range of state points and can thus be extrapolated to the critical colloid 
 packing fraction $\eta_{\text{col}}^{\text{crit}} = 0.103$ to determine 
 $\eta_{\text{pol}}^{\text{crit}} = 0.264(10)$. The inset shows the value of
 $W_{+}$($\propto \chi_{+}$) along the black dotted line. 
 The used colors correspond to the ones used in 
 the main legend. Note that the height of the maximum can not be extracted from 
 this analysis. However, we are only interested in the position of $\chi_{+}$, 
 which can be estimated to be inside the gray area of the inset.
 }
 \label{Fig4-homogeneous}
\end{figure}

We simulate state points on various paths that cross the rectilinear diameter
and determine the maximum of the susceptibilities on each of them. One exemplary
path is shown as the black dotted line in \autoref{Fig4-homogeneous}. The inset of the same 
figure shows the determined values of $W_{+}$ along this line and the region 
where the order parameter susceptibility reaches a maximum is colored in gray.
In order to extrapolate the susceptibility to the critical point, we are only 
interested in the position of this maximum, not the numerical value which 
would be needed in order to investigate the scaling behavior.
In the thermodynamic limit the susceptibility will diverge at the critical point.
Due to finite size effects this can not happen in our simulation, but the 
susceptibility will reach a maximum nonetheless. Therefore, plotting only the 
positions of the maximums on each path allows us to extrapolate them towards the critical
colloid packing fraction and thus find an approximation for the correct 
critical polymer packing fraction as shown in \autoref{Fig4-homogeneous}. 
The critical polymer packing fraction is determined to be 
$\eta_{\text{pol}}^{\text{crit}} = 0.264(10)$, which is slightly lower than the
value from the simulation along the rectilinear diameter.  

\subsection{Cumulant intersection for constant packing fractions}
\begin{figure}[!hbt]
 \centering
  \def\svgwidth{1\linewidth}
  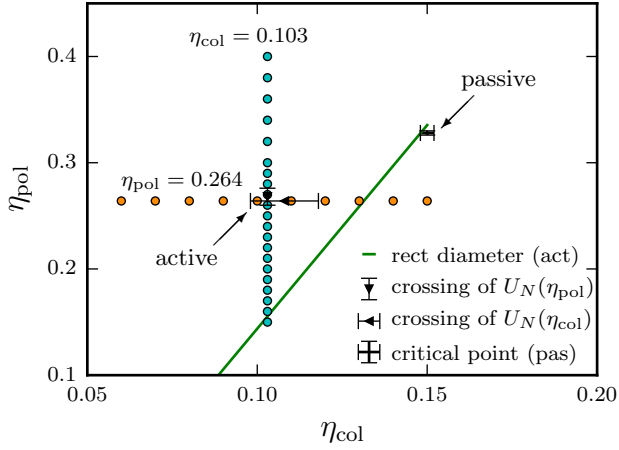
 \caption{%
 The simulated state points at constant colloid packing fraction (cyan) and 
 constant polymer packing fraction (orange) are shown together with the fitted 
 rectilinear diameter (\autoref{Fig2-phasediagram}) of the active system and the 
 critical point of the passive system taken from \cite{zausch2009statics}. 
 The intersections of both cumulants (\autoref{Fig6-const}) are shown as black 
 triangles. They overlap inside the error bars and clearly deviate from the 
 rectilinear diameter.
 }
 \label{Fig5-criticalpoint}
\end{figure}

As an alternative, we determine the critical packing fractions 
by simulating along a line in the parameter space that keeps one packing fraction 
constant. We use the result obtained in the previous section and simulate along 
a constant colloid packing fraction of $\eta_{\text{col}} = 0.103$ and a 
constant polymer packing fraction of $\eta_{\text{pol}} = 0.264$. The simulated
state points are shown in \autoref{Fig5-criticalpoint}.

\begin{figure}[!hbt]
 \centering
  \subfloat{\includegraphics{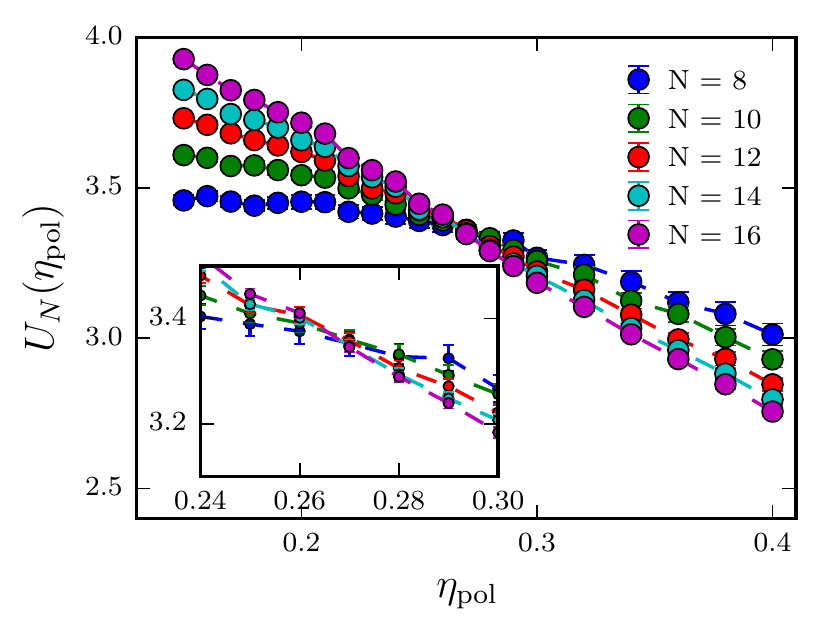}}\\\vspace{-0.9cm}
  \subfloat{\includegraphics{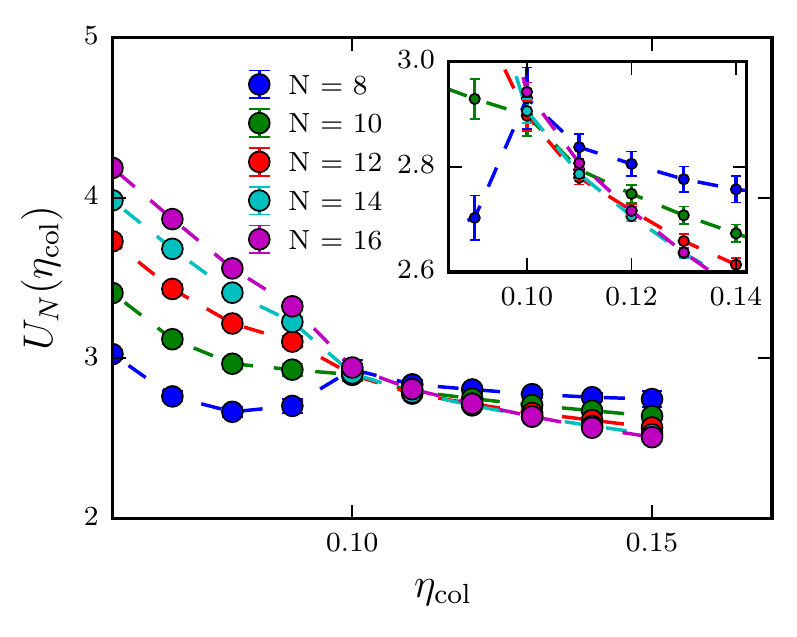}}\vspace{-0.5cm}
 \caption{%
 Crossing of the Binder cumulants along state points on a line of constant
 (a) colloid (b) polymer packing fraction as shown in \autoref{Fig5-criticalpoint}. 
 (a) The polymer cumulant $U_{N}(\eta_{\text{pol}})$ is plotted against the 
 polymer packing fraction $\eta_{\text{pol}}$ at constant 
 $\eta_{\text{col}}^{\text{crit}} = 0.103$. 
 The intersection point is read off as 
 $\eta_{\text{pol}}^{\text{crit}} = 0.268(8)$. 
 (b) The colloid cumulant $U_{N}(\eta_{\text{col}})$ is plotted against the 
 colloid packing fraction $\eta_{\text{col}}$ at constant 
 $\eta_{\text{pol}}^{\text{crit}} = 0.264$
 The intersection is at $\eta_{\text{col}}^{\text{crit}} = 0.108(10)$, % fill the correct intersection ! 
 which corresponds nicely with the previously determined intersection point.
 The inset in both figures shows the vicinity of the intersection point magnified.
 }
 \label{Fig6-const}
\end{figure}

From the intersection of the polymer cumulants at constant colloid packing 
fraction we determine the critical polymer packing fraction which should be in 
agreement with the polymer packing fraction that we determined before.
The run at constant polymer packing fraction is done to determine that the 
initial assumption was correct and the critical colloid packing fraction could 
be extracted from the simulation along the rectilinear diameter.
The results can be seen in \autoref{Fig6-const}. 
The cumulant intersection can be read off nicely and the crossing points agree 
within error margins.
For the critical polymer packing fraction we find 
$\eta_{\text{crit}}^{\text{pol}} = 0.268(8)$, which is in nice agreement with the
previously determined value of 
$\eta_{\text{extrapolated}}^{\text{pol}} = 0.264(10)$. The critical colloid
packing fraction is determined independently as 
$\eta_{\text{crit}}^{\text{col}} = 0.108(10)$.
The crossings from \autoref{Fig6-const}(a) and (b) are consistent with each 
other, which can also be seen in \autoref{Fig5-criticalpoint} since the error bars 
overlap.
As expected, the critical polymer packing fraction has to be slightly adjusted 
compared to the cumulant intersection from the rectilinear diameter, while the 
critical colloid packing fraction agrees within the margin of error.
As our best estimate for the critical point we choose the respective packing 
fractions obtained from the cumulant crossings at constant $\eta_{\text{pol}}$ 
and $\eta_{\text{col}}$ and thus obtain 
$\left(\eta_{\text{col}}^{\text{crit}} \middle/ \eta_{\text{pol}}^{\text{crit}} 
\right) = \left(0.108(10) \middle/ 0.268(8)\right)$.
Note that the intersection of the cumulants are to some degree insensitive to
minor variations of the other parameter as revealed by a comparison with the 
results from \autoref{secA}. 

\subsection{Critical exponent $\beta$}
With the knowledge of the critical point and the coexistence curve we can calculate the critical 
exponent $\beta$. The continuous Asakura-Oosawa model, which is used as a basis
for the active model discussed here, belongs to the Ising universality class \cite{vink2005}. 
In three dimensions one would thus expect $\beta = 0.3269(6)$ \cite{Talapov1996}.
Close to the critical point the magnetization $M$ scales in the Ising model as
\begin{equation}
    M = M_{0} \varepsilon^{\beta},
\end{equation}
with $\varepsilon$ being the distance to the critical point. 
In the continuous Asakura-Oosawa model this corresponds to:
\begin{align}
    M &= \sqrt{ \left( \eta_{\text{col}}^{\text{liquid}} - \eta_{\text{col}}^{\text{gas}} \right)^{2} + 
	        \left( \eta_{\text{pol}}^{\text{liquid}} - \eta_{\text{pol}}^{\text{gas}} \right)^{2}
	      } \\
    \varepsilon &=  \dfrac{
{\scriptstyle\sqrt{ 
      \left( \frac{1}{2} \big( \eta_{\text{col}}^{\text{liquid}} + \eta_{\text{col}}^{\text{gas}} \big) - \eta_{\text{col}}^{\text{crit}} \right)^{2} + 
      \left( \frac{1}{2} \big( \eta_{\text{pol}}^{\text{liquid}} + \eta_{\text{pol}}^{\text{gas}} \big) - \eta_{\text{pol}}^{\text{crit}} \right)^{2}
			      }
} 
			}{
{\scriptstyle
      \sqrt{\big(\eta_{\text{col}}^{\text{crit}}\big)^{2} + \big( \eta_{\text{pol}}^{\text{crit}}\big)^{2}}
			 }
}
\quad. \label{eq:eps}
\end{align}
\begin{figure}[!hbt]
 \centering
 \includegraphics{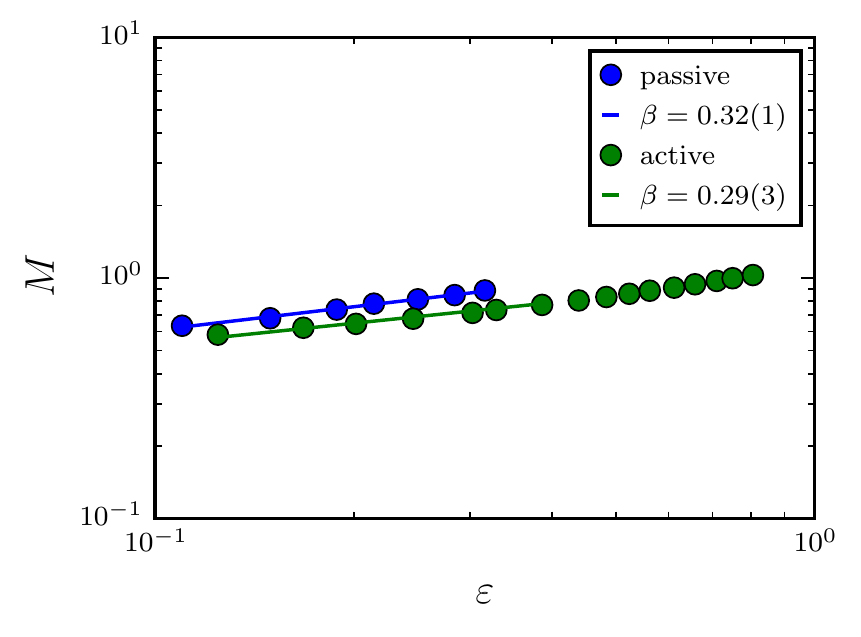}
 \caption{%
  Comparison of the critical exponent $\beta$ of the active and the
  passive model. The points are the respective values as extracted from the 
  phase diagrams in \autoref{Fig2-phasediagram}, 
  while the line represents a fit from which the critical 
  exponent $\beta$ is determined. The error is calculated by repeating the fit 
  for different critical points within the error bars.
 }
 \label{Fig7-beta}
\end{figure}
In \autoref{Fig7-beta} the order parameter $M$, calculated from the phase diagrams 
in \autoref{Fig2-phasediagram}, is plotted against $\varepsilon$ which was 
determined with the respective critical points (for the passive case we use the
literature value from \cite{zausch2009statics}) in a log-log plot.
For the passive case we recover the 3d-Ising value $\beta = 0.32(1)$ as expected.
The active case has a value of $\beta = 0.29(3)$ and is thus close to the 
value of the 3d-Ising universality class as well. 
Even though the fit in \autoref{Fig7-beta} matches the data points nicely, 
we have to assign a large uncertainty to the critical exponent for the active 
system. This is due to the error bars of the critical point, 
which in turn affects the estimation of $\varepsilon$ (\autoref{eq:eps}). 
In order to account for that, we have calculated $\varepsilon$ for various 
choices for the critical point (within the error bars) and repeated the fit. 
We then choose the error for $\beta$ as the standard deviation of all possible 
choices. For bigger $\varepsilon$ we get a deviation from the linear behavior 
on the log-log plot, 
and we thus do not account for them in the fit. 
The system, however, is only expected to follow this power law for small 
$\varepsilon$ anyway. It should be noted that the model discussed here will be
strongly influenced by the underlying passive model and thus one would expect to
find a crossover region between Ising critical behavior and possibly a 
critical behavior corresponding to the universality class for 
active particles. Other models for active particles that introduce a phase 
separation instead of facilitating an already existing one might be better 
suited to study the question of universality.

\subsection{Critical exponent $\nu$}
To determine the critical exponent $\nu$ we use the cumulant intersection of the 
polymers. The slope of the cumulants at the critical point can be extracted from 
\autoref{Fig6-const}(a). It is expected that $\frac{dU_{L}}{d\eta_{\text{col}}}$ 
scales with $L$ as \cite{vink2006critical} 
\begin{align}
  \dfrac{dU_{L}}{d\eta_{\text{pol}}} \propto L^{\frac{1}{\nu}} \quad.
  \label{eq:nu_vs_cumulant}
\end{align}
The slope at the critical point does not change rapidly, thus we determine it
via a linear fit over the five state points that are closest to the critical 
polymer packing fraction. 
However, the slope in \autoref{Fig6-const}(a) is negative, therefore we 
investigate the inverse cumulant $Q_{L}$ and rewrite \autoref{eq:nu_vs_cumulant} 
to 
\begin{align}
  \dfrac{dQ_{L}}{d\eta_{\text{pol}}} &\propto L^{-\frac{1}{\nu}} \label{eq:invcum}\quad.
\end{align}
\begin{figure}[!hbt]
 \centering
 \includegraphics{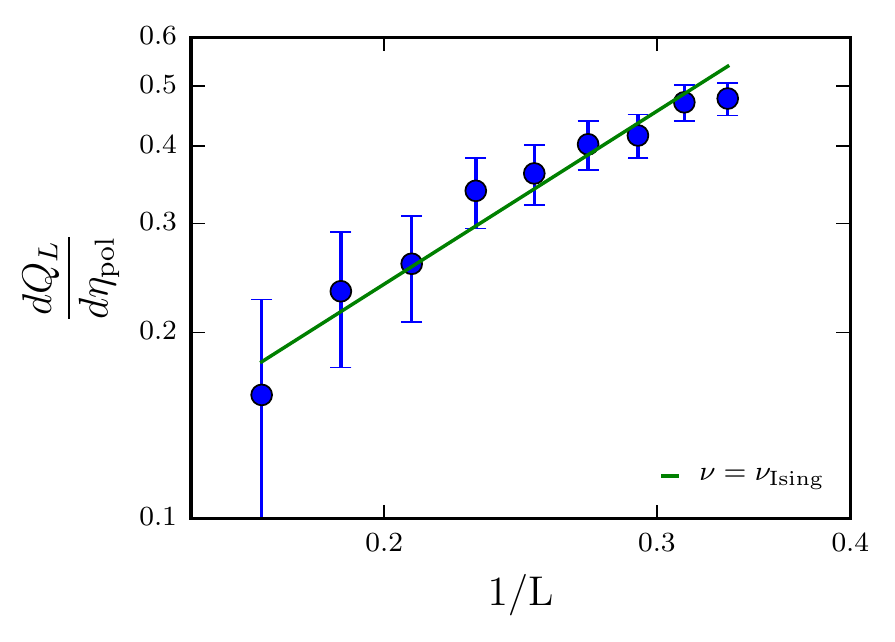}
 \caption{%
  Comparison of the critical exponent $\nu$ from the 
  slope of the cumulants from \autoref{Fig6-const}(a) 
  with $\nu_{\text{Ising}}$ \cite{hasenbusch2010finite}. 
  All integer values between $N=8$ and $N=16$ are considered in this figure.
  Both axes have a logarithmic scale.
 }
 \label{Fig8-nu}
\end{figure}
%The critical exponent $\nu$ can then be determined by a fit for all subboxes in
%the range of $N=8-16$.
The critical exponent $\nu$ is then compared to the 3d-Ising value of 
$\nu_{\text{Ising}} = 0.63002(10)$ \cite{hasenbusch2010finite} for all subboxes 
in the range of $N=8-16$, which is shown in \autoref{Fig8-nu}.
%We find $\nu = 0.64(6)$, which is in agreement with the $3$D Ising value of 
%$\nu_{\text{Ising}} = 0.63002(10)$ \cite{hasenbusch2010finite}.
While we get consistent results with the Ising value of $\nu$, the error bars of
the subsystems are large and the data range is very limited due to the limited 
range of subbox sizes so that the scaling is observed on less than a decade. 
This causes a large uncertainty in a fit to the data points in \autoref{Fig8-nu}
with \autoref{eq:invcum}, which results in $\nu = 0.64(6)$. 
For still smaller systems a plateau is expected as one can no longer observe any
fluctuations. For bigger systems correlations due to the finite size of the 
simulated box $S$ influence the system and the statistical accuracy is decreased.

\section{Discussion and Conclusion}
We have discussed a method of how to estimate the location of the critical 
point in a system of active particles analyzing the density fluctuations 
in subboxes. 
%The values for $\eta_{\text{pol}}^{\text{crit}}$ we get from both methods are 
%consistent with each other. 
The problem is difficult since one has to search in a two-dimensional space
of densities $\left(\eta_{\text{col}, \eta_{\text{pol}}}\right)$, and thus
the critical point of our active system can only be 
determined with modest accuracy as 
$\eta_{\text{crit}}^{\text{col}} = 0.108(10)$ and 
$\eta_{\text{crit}}^{\text{pol}} = 0.268(8)$. 
Note that the subblock-density-distribution method we used is general and as 
such should apply to each density driven phase separation.
The iterative approach that we have used to find the critical point is necessary since in this
model the order parameter of the phase transition is an a priori unknown linear
combination of both packing fractions. In a model with an intensive control
parameter, e.g. the temperature in a Lennard-Jones system, or the active
velocity in an active Brownian particle system, the search for the critical 
point is simpler.

The model used was chosen to feature a phase transition in the limit of no 
activity in order to have a critical point. While this model therefore is 
suitable to discuss the determination of the critical point it will be 
influenced by the underlying passive model. 
Our results for the critical exponents $\beta$ and $\nu$
are consistent with the Ising universality class.
For smaller values of $f_{A}$ we expect the critical point to steadily 
shift towards the passive value. 
%While the critical exponent $\beta$ will remain mostly unchanged.
%
If the active system very close to its critical point exhibits critical 
behavior of a different universality class, further away from the critical point
this is expected to be hidden by crossover effects. We can not rule out that 
this consideration is the correct interpretation of our findings.
%We have shown that the determination 
%of critical exponents is possible, but note that other models might be more 
%suitable to study the question of universality.

\section*{Acknowledgements}
We would like to thank S.K. Das, S. Egorov and M.P. Allen for fruitful discussions.
BT acknowledges the Graduate School Materials Science in Mainz for partial 
financial support in form of a DFG-fellowship through the Excellence Initiative
(GSC 266) as well as the SFB-TRR 146. 
BT, JS, TS and PV acknowledge support by the SPP1726 ``Microswimmers'' (grant 
number SP 1382/3-1 and VI 237/5-1). 
We thank the ZDV Mainz for computational resources.

%\clearpage{}

\bibliography{references}

\end{document}